\documentclass[printer]{aa} 
\usepackage{graphicx}
\usepackage{amsmath}
\usepackage{txfonts}
\usepackage{csquotes}
\MakeOuterQuote{"}

\begin{document} 

\title{Galactic Cosmic Rays measured by UVS on Voyager 1 and the end of the modulation:}
\subtitle{Is the upwind heliopause a collapsed charge-exchange layer?}

\author{R. Lallement
          \inst{1}
         \and
         J.L. Bertaux
         \inst{2}
          \and
          E. Qu\'{e}merais
          \inst{2}
          \and 
          B.R. Sandel\inst{3}
          }

   \institute{GEPI Observatoire de Paris, CNRS, Universit\'e Paris Diderot, Place Jules Janssen  92190 Meudon, France\\
              \email{rosine.lallement@obspm.fr}
        \and
        LATMOS, Universit\'e de Versailles Saint Quentin, INSU/CNRS, 11 Bd D' Alembert, 78200 Guyancourt, France
         \and
             Lunar and Planetary Laboratory, University of Arizona, Tucson, USA\\
             }

   \date{Received ; accepted }

 
  \abstract
 {The detectors of the UltraViolet Spectrographs (UVS) on Voyager 1/2 are recording a background intensity that was earlier assigned mainly to disintegrations in the radio-isotope thermoelectric generator and systematically subtracted from the signal to infer photon counting. Here we show that it arises instead from Galactic Cosmic Rays (GCRs). We show the GCR flux measured by UVS on Voyager 1 from 1992 to August 2013 and, by comparing with data from the GCR dedicated detectors, we estimate the energy range responsible for this UVS signal, around 300 MeV, and the response of UVS to the GCR anisotropy. 
  After the abrupt jumps of May and August 2012 the count rate has been fluctuating only slightly around a constant value, but comparing with data from the Low Energy Charge Particle Experiment (LECP) and the Cosmic Ray Subsystem (CRS) shows that those small variations are only responses to a varying anisotropy and not to a flux change. Taking advantage of the similarity in energy range to one of the products of the CRS instrument suite, we use the ratio between the two independent signals as a proxy for the temporal evolution of the GCR spectral slope around the 300 MeV range. We show that this slope has remained unchanged since August 2012 and find strong evidence that it will no longer vary, implying the end of the heliospheric modulation at those energies and that Voyager 1 at this date is near or past the heliopause.
The origin of this unexpectedly narrow and stagnating inner heliosheath is still unclear, and we discuss the potential effects of low solar wind speed episodes and subsequent self-amplified charge-exchange with interstellar neutrals, as a source of deceleration and collapse. We suggest that the quasi-static region encountered by Voyager 1 may be related to such effects, triggered by the strong post solar-maximum variability. This did not happen for Voyager 2 due to its trajectory at an angle further from the heliosphere axis and a later termination shock crossing. The existence on the upwind side of a mixing layer formed by charge transfer instead of a pure plasma contact discontinuity could explain various Voyager 1 observations.}


   \maketitle
%

\section{Introduction}
The two Voyager spacecraft are providing unprecedented data on the nature, structure, and size of the heliospheric boundary, measurements that are of fundamental interest for our understanding of the interplay between stars and interstellar matter in general, in the Milky Way and distant galaxies. Unexpected results on the solar wind termination shock (TS) and heliosheath properties have been obtained so far, and in August  2012 Voyager 1 entered a different region whose properties are very surprising and have given rise to new, fascinating controversies \citep{stone13,krimigis13,burlaga13,webber13}. Favoring a heliopause crossing and entry into the circumsolar interstellar medium (ISM) are the disappearance of the heliospheric energetic particles and the abrupt increase of the galactic cosmic rays followed by one year of stable flux. On the other hand, the direction of the magnetic field  has remained unchanged across this boundary and is still along the spiral heliospheric field, suggesting that the spacecraft is still cruising within the solar wind. Very recently \cite{gurnett13} have detected plasma oscillations whose frequency implies a density of 0.06-0.08 cm$^{-3}$, far above the solar wind heliosheath range, and instead very close to the expected circumsolar interstellar medium value deduced from the neutral hydrogen deceleration at the entrance in the heliosphere (0.04-0.07 cm$^{-3}$, \cite{izmod99}). It is not clear yet whether the steep density ramp inferred by \cite{gurnett13} between October, 2012 and April, 2013 corresponds to the heliopause discontinuity itself or the wider and more distant density enhancement in the pile-up region (e.g. \cite{malama06,zank13}), but in any case such data is a confirmation of the entry into the ISM, contrary to most predictions of a larger distance to the heliopause. Because obviously processes at play are far from being understood and require new models, and since no other in situ data will be available for decades, every piece of information is precious. The UltraViolet Spectrograph (UVS) \citep{broad77} has been used since the 1990's to measure the hydrogen Lyman-alpha "glow", the fraction of solar UV radiation that is scattered by inflowing interstellar atoms, at the periphery of the heliosphere. It has detected the emission of the so-called \textit{hydrogen wall}, the region of enhanced neutral H density that extends beyond the termination shock and is the neutral counterpart of the interstellar plasma  pile-up region in front of the heliosphere  \citep{quem96,malama06}. While such data provide their own constraints on the outer heliosphere structure, the UVS was not expected to provide any additional useful measurement related to this structure. Recently, a fraction of the UVS dataset has been used to identify the Milky Way H Lyman-alpha diffuse emission \citep{lall11}, a weak additional signal, and this identification has required carefully removing the background noise in the UVS detector. This contaminating signal was originally attributed primarily to gamma rays from the spacecraft radio-isotope thermoelectric generator (RTG). As a matter of fact, within 5 AU from Sun this signal was very slightly correlated with the solar flares, and for this reason a potential role of energetic particles, both solar and galactic, was considered as minor \citep{broad79}. However, revisiting this background signal revealed that it has increased since at least 1992, while at this time and later solar particles should no longer have any effect and since the RTG radiation has ineluctably slowly decreased (the half lifetime for Pu 238 is 88 yrs). Here we analyze in detail the Voyager 1 UVS background noise since 1993, and show that it is entirely or almost entirely due to the impact of galactic cosmic rays. Section 1 examines the entire V1 dataset and compares the temporal evolution of the extracted background noise with  galactic cosmic rays (GCRs) fluxes directly measured with dedicated Voyager instruments, to establish its spectral response. Section 2 discusses the sensitivity of the UVS detector to particle velocity directions and makes use of comparisons with published Voyager 1 data to assess it. Section 3 shows how both the  flux of GCRs detected by UVS and its spectral slope are constant since August 2012, in agreement with other Voyager 1 measurements. Section 4 discusses the deceleration of the heliosheath flow under the unique action of charge transfer with interstellar neutrals, including the gyromotion of the ions, and its potential role in the observed stagnation region and proximity of the heliopause.

   \begin{figure*}
   \centering
\includegraphics[width=0.9\linewidth]{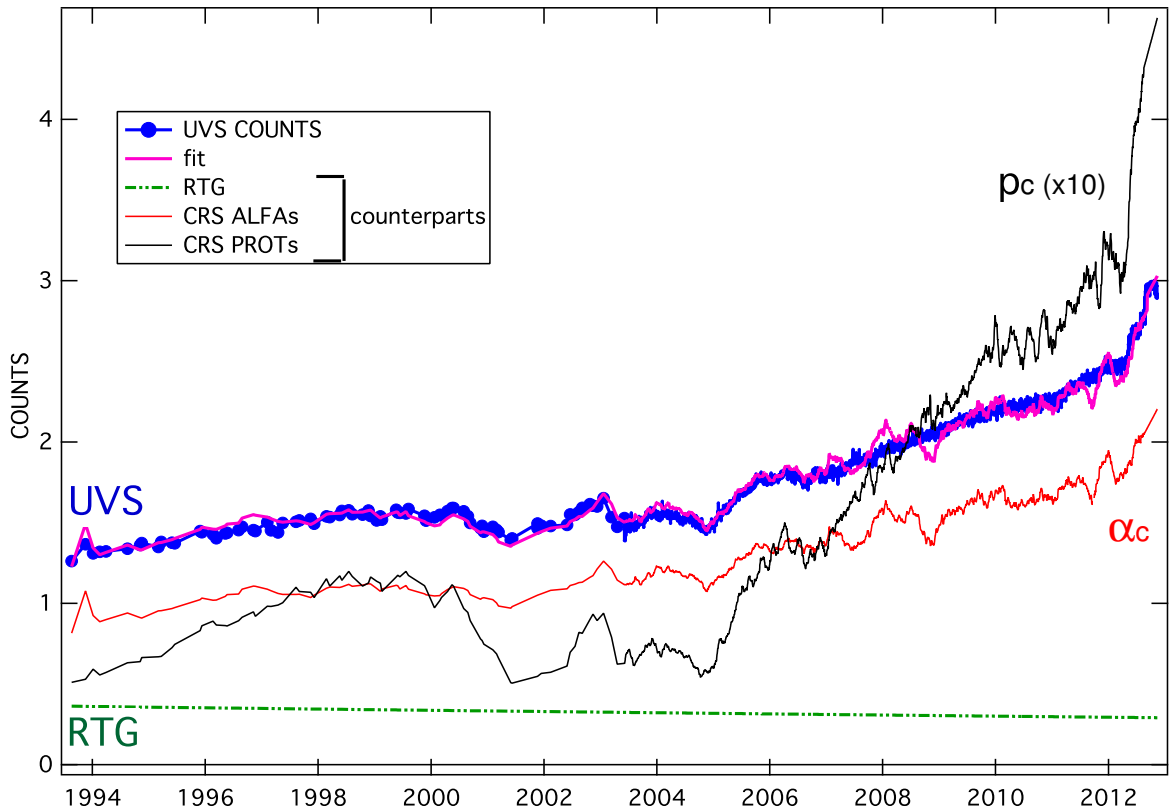}
   \caption{First estimate of the energy response of the UVS detector: the UVS count rate is displayed in blue. It has been fitted by a linear combination of two fluxes directly measured by the CRS instruments, the 133-242 MeV proton channel (black curve) and the 242-476 MeV alpha channel (red curve) plus an exponentially decreasing signal with half-life of 88 yrs assigned to RTG. Both the solar maximum (1999-2003) modulation amplitude and the average 1993-2012 gradient of the UVS signal are slightly stronger than those of the alpha particle parameters, and significantly smaller than for the protons, showing that UVS is sensitive to energies only slightly smaller than the alpha energy range. The coefficients are 91\% and 9\% respectively for the higher (as in the alpha channel) and lower (proton channel). In the graph the lower energy contribution has been increased by a factor 10 for clarity. The RTG contribution is minor.}
              \label{Figenergies}%
    \end{figure*}
   \begin{figure*}
   \centering
\includegraphics[width=0.9\linewidth]{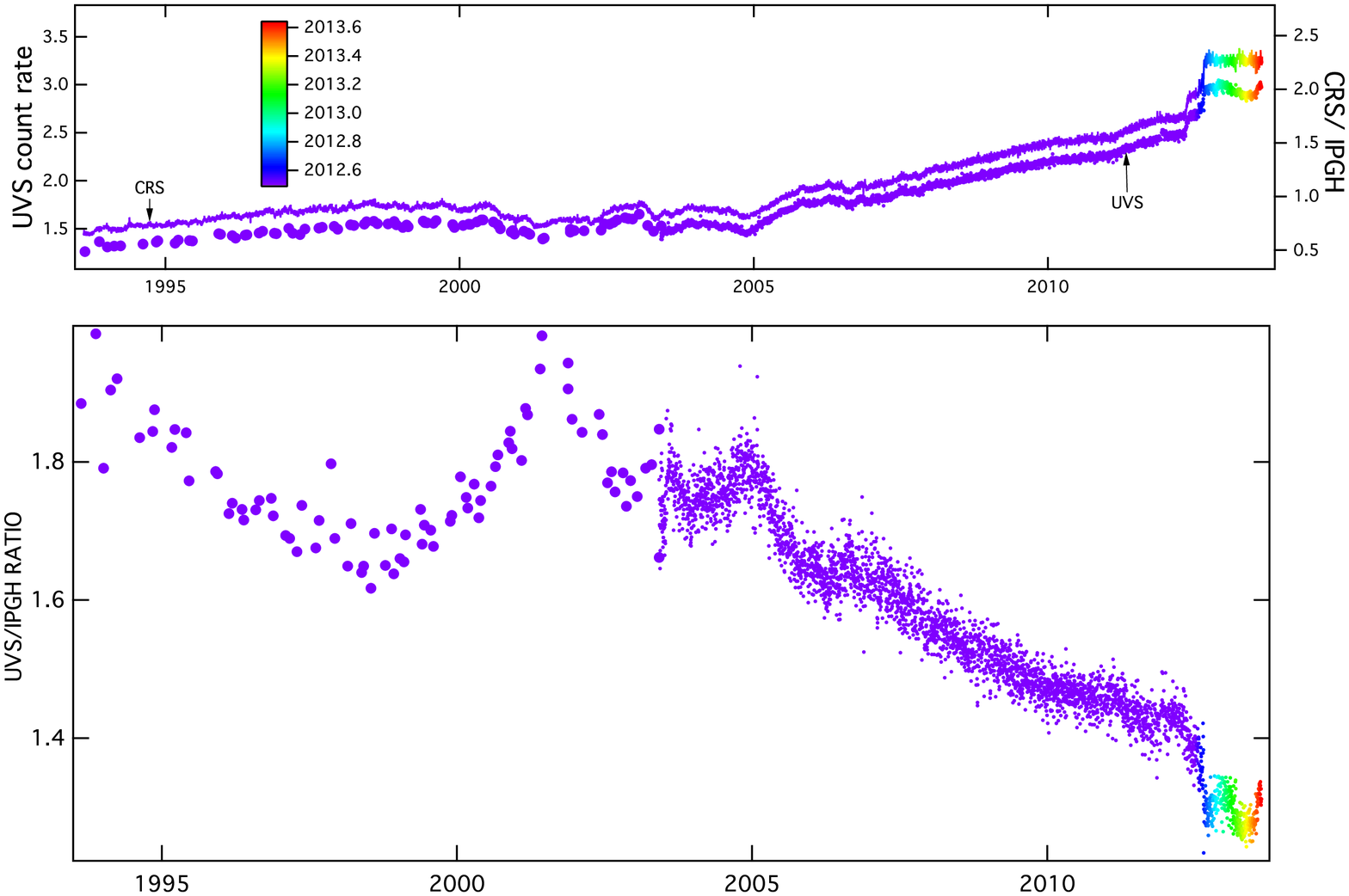}
   \caption{Comparison between the UVS background signal and the CRS count rate for protons above 70 MeV (hereafter IPGH), since 1992 (top). The color scale refers to time, and serves to separate pre- and post-jumps data and to allow comparisons with Fig. \ref{Figdemodall}. The two signals, although being fully independent measurements, are strikingly similar, allowing informative comparisons. Their ratio is displayed as a function of time (bottom panel). The global decrease and the temporary increases around the 1999-2003 solar maximum show that the energy range for UVS is globally higher than for the IPGH channel.}
              \label{Figratio}%
    \end{figure*}

\section{The UVS detector background and its sensitivity to GCRs}
\subsection{The UVS detector: configuration and background}

In the UVS detector, a pair of microchannel plates (MCP)  in series is placed at the focus of an objective grating spectrograph. A single photo-electron created at the input of the MCP generates a pulse of many electrons at the output, and this charge is collected on a linear array of 128 elongated anodes (channels) that correspond to the 540 to 1700 \AA\  range. The field-of-view is a 0.1$^{\circ}$ x 0.87$^{\circ}$ rectangle. The Ly$\alpha$ emission is inferred by integrating over the 9 UVS spectral channels over which the emission line is spread. Lines-of-sight close to nearby hot stars are avoided.  

The whole detector was protected from high energy particles of the Jovian radiation belts by a thick copper shield, open in the direction of light coming from the grating. Such a configuration explains why the whole chain of high energy particle detection is not fully isotropic, as was shown by the detailed study of the background behavior during scan platform motions \citep{lall11}. The background spectrum has been described in the Supporting Online Material of \cite{lall11}. Here we have extracted the cleanest signal by summing the count rate in all channels excepted those that can be contaminated by heliospheric and Galactic UV radiation (i.e. the Ly$\alpha$ H glow, bright early-type stars and Milky Way emission), the first and last ones whose behavior may depend on the instrument mode, and finally three 3 \textit{hot} anodes that started to behave differently in December 2011 when the heating of the instrument was reduced. We have extracted the sum of the count rates in the 59 selected channels from both the scans that were performed between 1992 and 2004 and the more recent measurements in a fixed direction \citep{quem95,quem08,lall11}. A unique position of the scan platform supporting UVS was used for the fixed direction data, and to ensure homogeneity of the directional response we restricted the earlier, scan data to those recorded at the same, unique position.   Integration times for all data are of the order of  10 hours. It is difficult to assign a measurement error to this background based on the integration time and the number of channels, since we do not know the number of channels impacted simultaneously by a single particle. We will see below from comparisons with other data that the observed variance of the signal is less than 1\%, but a fraction of it or all of it it may be the actual variance of the measured quantity, i.e. a particle flux variability. During the scan period from 1992 to 2004, the background has been extracted at intervals of $\simeq$ 1.4 month, while since 2004 data are recorded on a quasi-daily basis.

\subsection{Temporal evolution of the UVS background}
This UVS detector background is displayed as a function of time in Fig \ref{Figenergies}.  It is characterized by a quasi-monotonic increase, except for drops between 1999 and 2005. Clearly it can not be due to the radio-isotope thermoelectric generator (RTG), because that source is expected to decrease exponentially. Since 1999-2005 was a time of strong solar activity, obviously the background is directly or indirectly linked to this activity, with a negative correlation suggestive of a modulation of GCRs by the heliosphere. On the other hand,  it  can be seen that it is very similar in shape to the direct particle measurements of GCR protons and alphas by the CRS instrument, displayed in the same Figure (data available from the http://voyager.gsfc.nasa.gov website). This leaves no doubt that the origin of the UVS background signal is dominated by GCR's. In contrast to dedicated Voyager particle instruments, the UVS detector has no energy resolution, but it is possible to infer the energy range that is mostly contributing to the count rate by comparing this overall temporal evolution with that of data from other instruments. Such an indirect calibration is based on the fact that, in a stationary heliosphere, the higher the energy of the GCR, the easier its entrance into the heliosheath and supersonic solar wind regions. Models of GCR propagation that include diffusion, drifts and convection all predict such a general behavior \citep{ipaxford85,jokipii93,webber09}. Consequently, for a S/C leaving the heliosphere, the higher the particle energy, the smaller the flux increase. It is thus possible to obtain a first estimate of the energy range based on the global intensity gradient. On the other hand, in non-stationary conditions, in particular during marked increases of solar activity, particles of different energies do react differently to the associated solar wind and magnetic field perturbations. Because particles of high energy are less affected than less energetic ones, the relative depth of the flux decreases around solar maxima provides a second estimate of the energy range of the detected particles. By examining  publicly available dedicated data, we found that the temporal evolution of the UVS background signal is the closest in shape to that of the 192-476 MeV alpha particle fluxes measured by CRS displayed in Fig. \ref{Figenergies}. More precisely, the temporal gradient and the relative amplitude of the large modulation linked to the solar cycle 23 maximum are very similar. Guided by this first estimate we successfully modeled the UVS background signal as the sum of two distinct components, the first proportional to the alpha channel signal just quoted (for 91\%), and the second proportional to the CRS 133-242 MeV proton channel (for 9\%), plus a signal exponentially decreasing with an 88-yr half-life to represent the RTG contribution and representing less than 8\% of the total signal in 2012 (Fig. \ref{Figenergies}). The closest similarity to the alpha channel signal reflects only the particle energy range and does not mean that UVS is more sensitive to alphas: since there is no mass nor charge selection, and GCR protons are far more numerous than $\alpha$ particles, most of the UVS signal is certainly due to protons. The sensitivity to this high energy range is very likely due to the thick shield of copper that protected the UVS detector from radiation belts during Jupiter encounter.

\section{Comparison with CRS data and temporal evolution of the GCR spectral gradient around 300 MeV}
\begin{figure*}
 \centering
\includegraphics[width=0.9\linewidth]{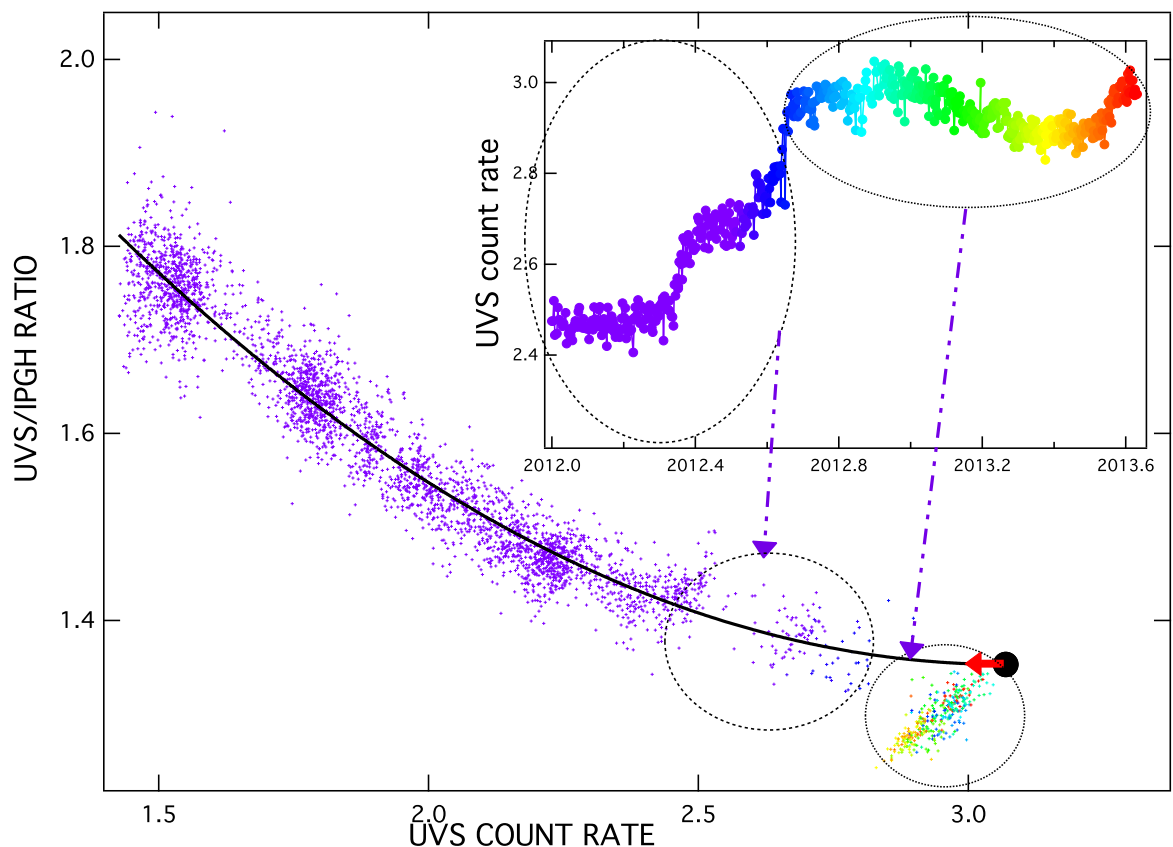}
   \caption{The main curve shows the UVS-IPGH ratio, this time as a function of the UVS signal. All time periods are now merged to form a smooth decreasing curve that prevails until the latest intensity jumps. Its slope (absolute value) is also decreasing and the flattening is already pronounced in 2012.   The large black dot corresponds to the extrapolated minimum of this differential demodulation curve, fitted by a third order polynomial and based on data prior to 2012.6 (in violet). It shows that the ratio achieved in 2012.6 is very close to this extrapolated minimum.
   The inset shows UVS measurements from May 2012. The color scale marks the time since 2012.5, and allows to find the corresponding data points in the main curve. From this main curve it is clear that after the second jump the UVS/IPGH ratio now significantly departs from the global trend that was prevailing since 1992. 
   The small fluctuations of the UVS signal since Sept 2012 (inset) correspond to UVS-IPGH varying ratios going up and down around an average value, that can be also followed in the main curve. As will be demonstrated using data from the LECP instrument, this behavior is linked to post-jump varying anisotropies of the proton distribution.}
              \label{Figdemodall}%
    \end{figure*}

We have then compared in more detail the UVS data with data from one of the CRS High Energy Telescope measured count rates, shortly named IPGH, that is stacking all detections of protons above 70 MeV. CRS IPGH is available on-line,  provides data with high statistical significance and very high temporal resolution, with between two and four measurements per day. Despite their total independence, the similarity between the IPGH and the UVS rates is striking, as shown in Fig \ref{Figratio}. Indeed, fitting this IPGH rate to the same two CRS energy-selected channels, in the same way it was performed for UVS leads to a similarly good adjustment, with 79 \% and 21\% for the higher and lower energy channels respectively. This suggests that the IPGH energy range is only slightly lower than that of UVS. Such a small difference is confirmed by the ratio between the signals from the separate instruments, the UVS and the CRS IPGH, also shown in Fig \ref{Figratio}. As a matter of fact, the UVS/IPGH ratio is globally decreasing since 1992, and the relative amplitude of the solar maximum decrease is smaller for UVS. Both trends confirm the small shift towards higher particle energies.
Their good statistics, strikingly similar temporal evolutions and the proximity in their energy ranges makes the two CRS IPGH and UVS signals well suited for studying the energy differential modulation of GCR protons around the $\approx$ 300 MeV range. Fig \ref{Figdemodall} displays the UVS/IPGH ratio as a function of the UVS signal. For the entire period from 1993 to mid-2012 all data points, including the 1999-2003 measurements, are now aligned to form a smooth curve. This UVS/IPGH ratio can be taken as a proxy to the slope of the modulated proton energy spectrum at $\approx$ 300 MeV, while the recorded intensity (by CRS or UVS) at time t (or equivalently the particle flux) can be considered as a proxy for the demodulation at this time t. Therefore the curve represents the evolution of the spectral slope as a function of the level of demodulation. According to the modulated spectra that are based on the most sophisticated propagation models (e.g. \cite{ipaxford85,jokipii93,webber09,sherer11,herbst12,fisk12}) and neglecting non-stationary effects, such flux ratios of adjacent energy ranges (here in the sense of higher to lower energy) can only decrease when moving away from the sun. Further, they should reach a constant value at the same time the fluxes reach the unmodulated levels (outside of the heliopause if the outer heliosheath effects are negligible, farther away in the opposite case \citep{sherer11,herbst12}), whatever the actual shape of the unmodulated spectrum. Interestingly, the negative gradient of the ratio is quite small for data recorded since 2011, as can be seen using the color codes for the dates. This flattening, associated with the fact that the gradient can only be negative (or null), suggests that the proton spectra at energies of the order of 300 MeV are not far from being entirely demodulated from the heliosphere and inner heliosheath effects. The potential (but not necessary) remaining demodulation can be estimated by extrapolating the curve in Fig \ref{Figdemodall}, again excluding data after June 2012, down to a minimum. We have tentatively drawn this potential limit, shown in Fig \ref{Figdemodall} at the extremity (defined where the derivative is 0) of a polynomial fit of order 3 to the data. 

As was mentioned, the UVS background may contain a minor contribution due to the RTG. The above study of the UVS/IPGH ratio did not take into account such a contribution, and we must consider its potential effect. The RTG signal is by intrinsic nature perfectly smooth, and due to the very long decay time, almost constant over the time period we considered. As a  consequence its only potential effect is a slight increase of the temporal variation of the UVS/IPGH ratio w.r.t. to the one we have inferred within the assumption of a null contribution. We have investigated the effects of a potential contribution like the one from Fig \ref{Figenergies} and found no difference in the conclusions. 

In Fig \ref{Figdemodall} all data points corresponding to the post-June 2012 period fall significantly below the main curve we have discussed in this section, marking a clear change from the general trend. However, as we will demonstrate below, this discontinuity has a source other than a change in the differential modulation. 
   \begin{figure*}
   \centering
\includegraphics[width=\linewidth,height=11cm]{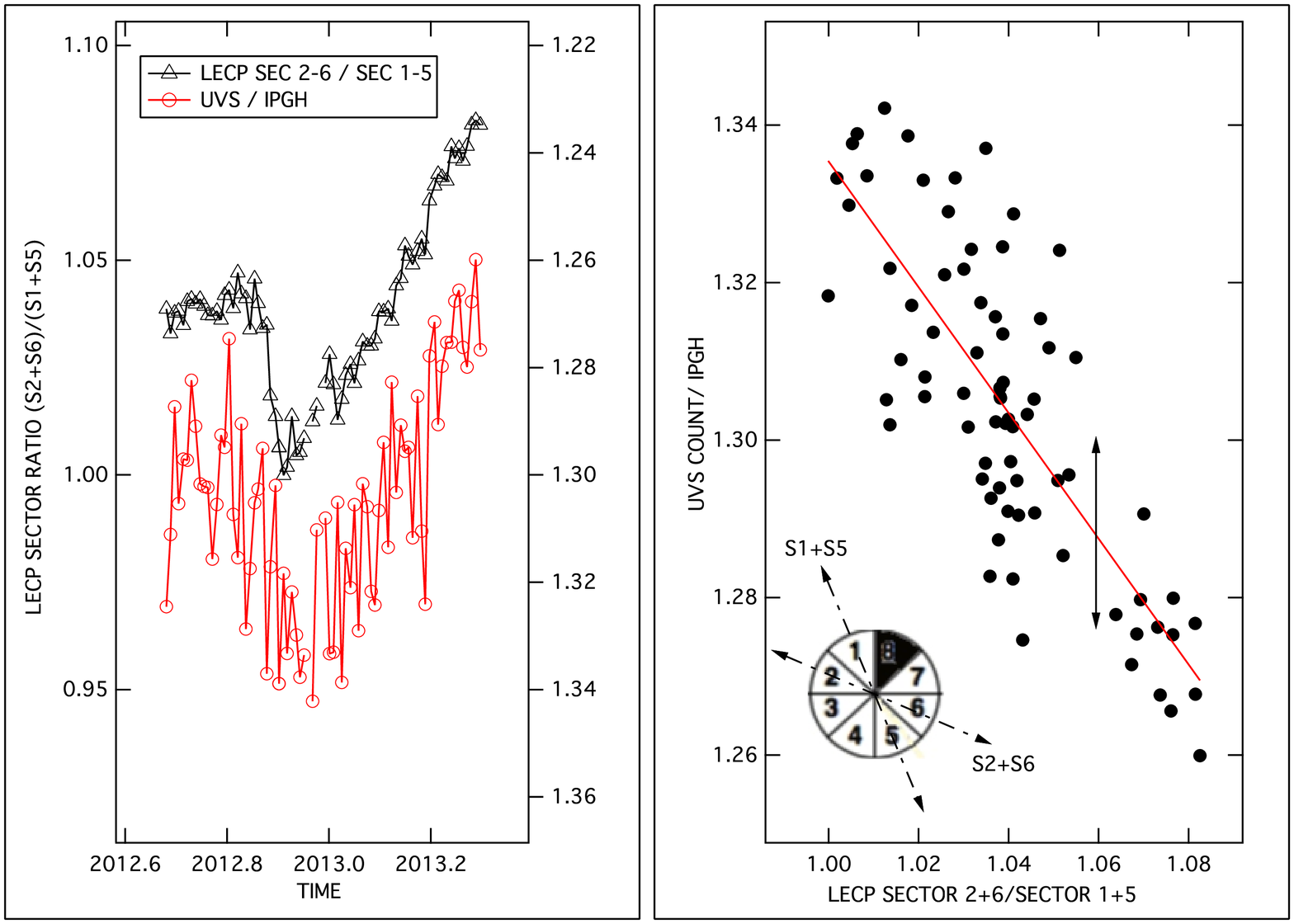}

   \caption{(Left) Comparisons between the LECP sector-based measured anisotropy \citep{krimigis13} between 2012.63 and 2013.25, and the UVS-IPGH ratio at the same period. There is a strong correlation between the two quantities. The linear relationship adjusted to the data (right) will be used to correct for the flux anisotropy during this period of time. The black arrow represents the variance of the UVS-IPGH ratio.}
 \label{Figcalib}%
\end{figure*}
  \begin{figure*}
 \centering
\includegraphics[width=\linewidth]{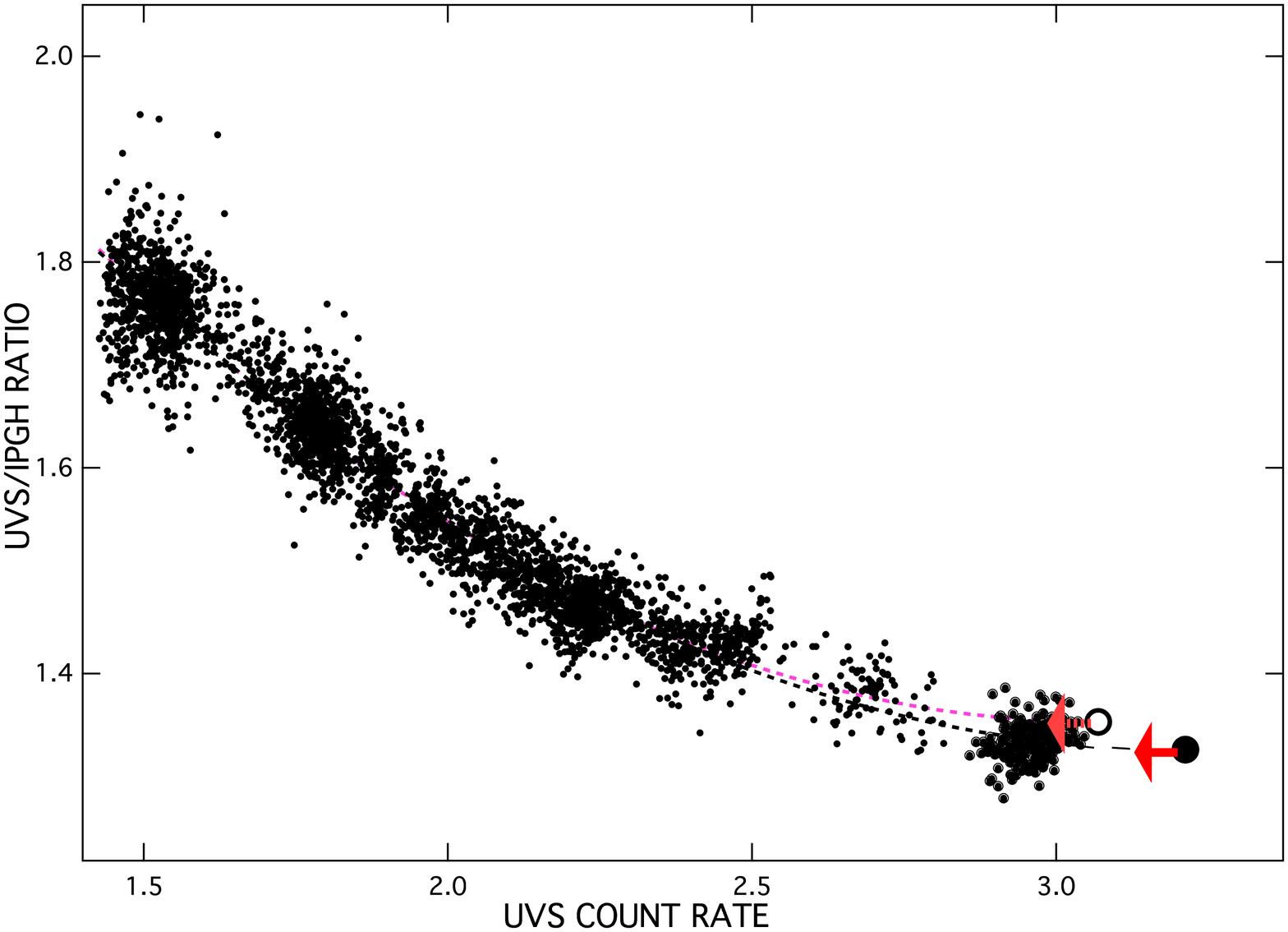}
   \caption{Same as Fig \ref{Figdemodall} (main curve), excluding  post-jumps data except those being corrected for anisotropies, based on Fig \ref{Figcalib}. The discontinuity has now disappeared. Hypothetical minima deduced from the polynomial fit of Fig \ref{Figdemodall} and from a new one one including the post-jump corrected  data are marked by filled and empty circles respectively  (see text).}
              \label{Figdemodcorrec}%
    \end{figure*}

\section{Latest data: the May and September jumps}
We now discuss in more detail the latest data. As can be seen in Fig \ref{Figdemodall}, since May 2012 the UVS has recorded two very prominent jumps that were similarly recorded by all GCR instruments, in May 7 and September 2012. There are spectacular differences between the two jumps as far as other data are concerned, the second one corresponding to the total disappearance of the heliospheric energetic particles \citep{stone13,krimigis13}. After the second jump the UVS signal did not stop to fluctuate with a small (few \%) amplitude around a stable post-jump value. There are two maxima separated by about 9 months, the second one having occurred recently in June 2013. Contrary to this pattern, the IPGH rate has remained at a constant average level. 
As a result of those different behaviors, the UVS/IPGH ratio is remarkably different and evolves remarkably differently after the last jump, in comparison with all earlier data. The ratio falls significantly below the value that can be extrapolated from the earlier data, and then fluctuates. Its evolution can be followed  in Fig \ref{Figdemodall}, using the color codes. During the two periods with maximum signal, the ratio is quite close to an extrapolation of the global demodulation trend, while during the minimum (end of May-June 2013, yellow points) the ratio is about 5\% smaller. As we said, given the IPGH constancy, those variations are entirely due to the UVS count rate variations. 

We now consider this behavior as a potential consequence of post-jump particle distribution anisotropies. As a matter of fact we know from earlier studies \citep{lall11} that the UVS detector chain of particle detection is not isotropic, due at least partially  to the shape of the copper shield mentioned above \citep{broad79}. The LECP instrument has provided crucial information on the particle distribution anisotropy; in particular it showed that a strong depletion in particles with 90$^{\circ}$ pitch-angle values prevailed for about two to three months after the second jump, disappeared, then reappeared, based on data  sorted by angular sectors \citep{krimigis13}. We have compared the UVS count rate and the ratio between two different LECP measurements of protons above 211 MeV, namely the combination of sectors 2 and 6 on one hand, that corresponds to a mean direction roughly parallel to the magnetic field, and the combination of sectors 1 and 5 on the other hand, that corresponds to a mean direction at about 60 degrees from the first one. The LECP sector data were taken from the figures of \cite{krimigis13}. The two signals are displayed in Fig. \ref{Figcalib}, and are found to vary in very similar ways. This clearly demonstrates that the post-jump anisotropic pitch-angle distribution is reflected in the UVS count rate, at variance with the CRS data. More quantitatively, we find that during the period of about 7 months following the second jump, the UVS/IPGH ratio is proportional to the LECP sector ratio defined above, providing a calibration of the UVS response to the anisotropy degree.  

We have made use of this calibration based on LECP sector data to correct the UVS rate for the prevailing anisotropy. To do so we simply multiplied the measured rate by a quantity proportional to the LECP sector ratio, assuming a reference value of 1 (no correction) when LECP measured a full isotropy, in November 2012, days 331-339. In doing so, we ignore potential varying anisotropies along an axis perpendicular to the plane probed by LECP. 
When removing post-jumps measurements that could not be corrected in this way and keeping the corrected one, the differential modulation curve of Fig \ref{Figdemodall} becomes the one drawn in Fig \ref{Figdemodcorrec}. It can be seen that the discontinuity that appeared after the jumps has now disappeared, suggesting that anisotropies are the major source of this discontinuity. As mentioned above, the flattening of the curve and the extrapolated minimum ratio constrain the end of the demodulation process, because the spectral slope must either remain constant or decrease as Voyager 1 is progressing. The same type of extrapolation as in Fig \ref{Figdemodall} can be done, this time using data both before and after the jumps.   It can be seen that the demodulation is either completed (in this case  the group of the latest points corresponds to the end) or is very close to completion (the end is somewhere between the latest points and the extrapolated minimum). In a quantitative way, the maximum possible additional increase in the UVS rate suggested by the polynomial fit is of the order of 5\%. We have to bear in mind that the correction made using LECP data is based on a solid angle smaller than 4$\pi$. It is possible that a better correction based on the full distribution would have  changed the results slightly. However,  this does not change our general conclusion that the demodulation has ended or was very close to its end in August 2012. This is in agreement with the results of \cite{webber13} and \cite{webber13b,webber13c} and the more recent measurements of a high density by \cite{gurnett13}. On the other hand, as already mentioned, a heliopause crossing is contradicted by 
the absence of magnetic field deviation \citep{burlaga13}, and the steep density increase of $\simeq$20\% per AU recently inferred by \cite{gurnett13}  is hardly compatible with the plasma pile-up region that comes out from MHD-kinetic models (e.g. \cite{izmod05,malama06,zank13}) and resembles more the expected sharper transition from the inner heliosheath to the ISM. 

\begin{figure}
\centering
\includegraphics[width=0.8\linewidth,height=7cm]{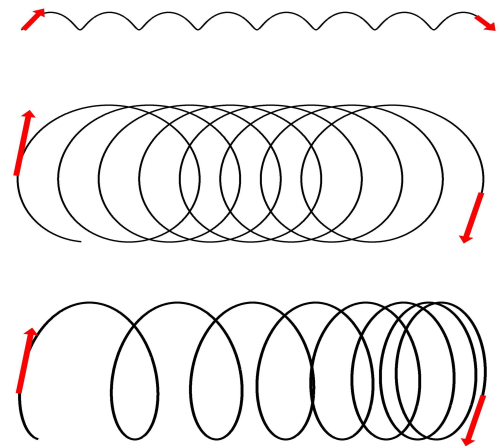}
   \caption{Top and middle: Schematic illustration of the differences between the thermal (top) and suprathermal (middle) ions regarding CE rates: for a given time interval the number of CE reactions for the two species are proportional to the lengths of the helicoidal curves followed by the ions. The cross-sections do not vary strongly from one species to the other, but the lengths directly depend on the individual speeds that are much higher for the suprathermal species, hence higher CE rates. Bottom: for all types of particles, the smaller the wind bulk speed, the higher the number of charge-exchange reactions, for the same distance traveled by the bulk flow. Because CE reactions with counter-streaming H atoms result in a deceleration of the flow, this constitutes a positive feedback mechanism.}
 \label{helices}%
\end{figure}

\section{Charge-exchange in the heliosheath}

Whether or not Voyager 1 has exited the solar wind in August 2012 at $\simeq$ 122 AU, it remains that the end of the 300 MeV GCR modulation discussed above, the recently observed GCR spectra, the GCR flux jumps and especially now the plasma oscillations, all observations show that the heliosheath is significantly narrower and the heliopause nearer than most global models predict (e.g. \cite{izmod05,malama06,romana08,zank09,alouani11,pogorelov08,zank13}). This suggests that some process or combination of processes may have been underestimated in global models of the heliospheric interface. Models \textbf{considering suprathermal ions either by means of a global kappa distribution (see \cite{heerik08}) or of an additional Maxwellian population \citep{malama06} }result in a narrower heliosheath, but they can not simultaneously explain the two other observed and unexpected characteristics of the inner heliosheath: -(i) the flow pattern along the V1 trajectory is strikingly different from the one at Voyager 2: as a matter of fact along the V2 trajectory there is no strong and early deceleration as is the case for V1, instead the flow is diverted toward higher latitudes. -(ii) the magnetic flux is conserved along the V2 trajectory,  while it is significantly decreasing for V1 \citep{krimigis11,richardson11,decker12,richardson13}. Those differences remain unexplained, even if solar cycle effects may be responsible for some of them \citep{pogorelov12}, and multiple reconnections for some magnetic flux variability \citep{swisdak13}. 
In this section we consider the influence on the ion-H atoms charge-exchange (CE) rates and on the subsequent flow deceleration, of the combination of several effects: \\
-the coexisting large individual speeds and low bulk speeds that characterize the inner heliosheath, \\
-the coexisting thermal and suprathermal populations, and\\
-the peculiar solar cycle circumstances that prevailed along the V1 trajectory after the TS crossing. \\
We suggest that unusually frequent  and self-amplified charge-exchange reactions of ions with interstellar H atoms may lead to rapid momentum loss and deceleration of the inner heliosheath solar wind and large mixing of the solar and interstellar plasma through CE reactions, and that the conditions required for those effects were fulfilled along the V1 trajectory.

Among all locations in the entire heliosphere, the inner heliosheath is characterized by the largest imbalance between charged and neutral particles, the latter population being by far denser, about 0.1-0.2 atoms compared to about 0.005  ions per cubic centimeter beyond the termination shock. As a consequence, this is where one should expect the strongest influence of the neutrals on the plasma. Moreover, this is where the CE rate is the most strongly amplified by the large individual ion velocities, larger or comparable to the bulk speed (We define the \textit{individual speed} as the ion speed in the bulk frame). The inner heliosheath is composed of two main ion populations, thermal and suprathermal. It is now consensually accepted that the pickup ions (PUI's), after they have been gradually generated in the supersonic solar wind and convected outward, are preferentially accelerated at the TS, where they capture the major part of the kinetic energy of the upstream flow (e.g. \cite{gloeckler05,zank96,fahr10,burrows10}) and constitute a suprathermal population. This preferential acceleration explains the relatively low temperature of the post-shock ambient wind, of the order of 1-2 10$^{5}$K along the V2 trajectory \citep{richardson08}. The suprathermal fraction of the heliosheath wind is thus very likely significant, namely on the order of 20\%, the fraction of PUI's in the supersonic solar wind upstream of the shock at V2 \citep{richardson08_decel}. For ions moving in a monokinetic neutral H \textbf{atom} gas the probability of experiencing a CE reaction with an atom is the number of atoms contained in the volume swept by a solid whose transverse area is the ion-neutral CE cross-section $\sigma$ and length is the distance traveled by the ion through the neutral gas, or equivalently the product of the volume density of neutrals, the cross-section and the traveled distance if the cross-section is constant and the neutral gas homogeneous.

\begin{equation}
\frac{dn}{n}=- \sigma * n(H) * dl = - \sigma* n(H) *  |V_{ion}-V_{H}| * dt
\end{equation}

where $|V_{ion}-V_{H}|$ is the relative velocity vector between the two species.  In the supersonic solar wind ion thermal speeds are much smaller than the bulk speed, except for the PUI's, and thus ion trajectories are close to straight lines. In this case one can consider only the bulk speed to calculate the CE rate, i.e. $V_{ion}=V_{bulk}$, which is what is often done. This is no longer the case if the individual ion speed is comparable to the bulk speed or higher, e.g. PUI's in the supersonic wind. Due to the gyromotion, the swept volume is the one of a helicoidal tube and for a given distance traveled by the bulk flow the total distance traveled by the particle is larger by a factor that depends on the ratio between the ion speed and this bulk flow speed. In the inner heliosheath this is the case not only for all supra-thermal  ions (former PUI's), but also for the ambient wind since the bulk speed is of the order or smaller than $\leq$ 100 km.s$^{-1}$ while the proton thermal speeds are of the order of 30-40 km.s$^{-1}$.   For a given distance traveled by the ion guiding center, the length of the helicoidal figure increases with the ion gyrospeed, and as a consequence the CE rate per unit time is NOT the same for thermal ions and suprathermal ions: it is significantly larger for suprathermal ions whose speed is much larger than the thermal speed. This is illustrated in Fig \ref{helices} that shows in a schematic way typical trajectories of the two species.

Sophisticated stationary fluid-kinetic or MHD-kinetic global models (e.g. \cite{alex05, malama06, heerik08, zank09}) that take into account the ion velocity distributions are also taking into account the large rates of CE reactions experienced by the high speed ions in the inner heliosheath, however they do not consider peculiar flow conditions far from the stationary case nor distribution anisotropies. On the other hand extensive non-stationary global models (e.g. \cite{pogorelov08,pogorelov12}) represent well the non-stationary heliosphere but do not include multiple ion populations. Our goal here is solely to draw the attention on the potential consequences of peculiarly low solar wind bulk speeds in the presence of the high individual ion speeds that are characteristic of the inner heliosheath, in particular on a potential self-amplification of the CE deceleration and of solar-interstellar mixing.



\begin{figure}
\centering
\includegraphics[width=0.95\linewidth,height=7cm]{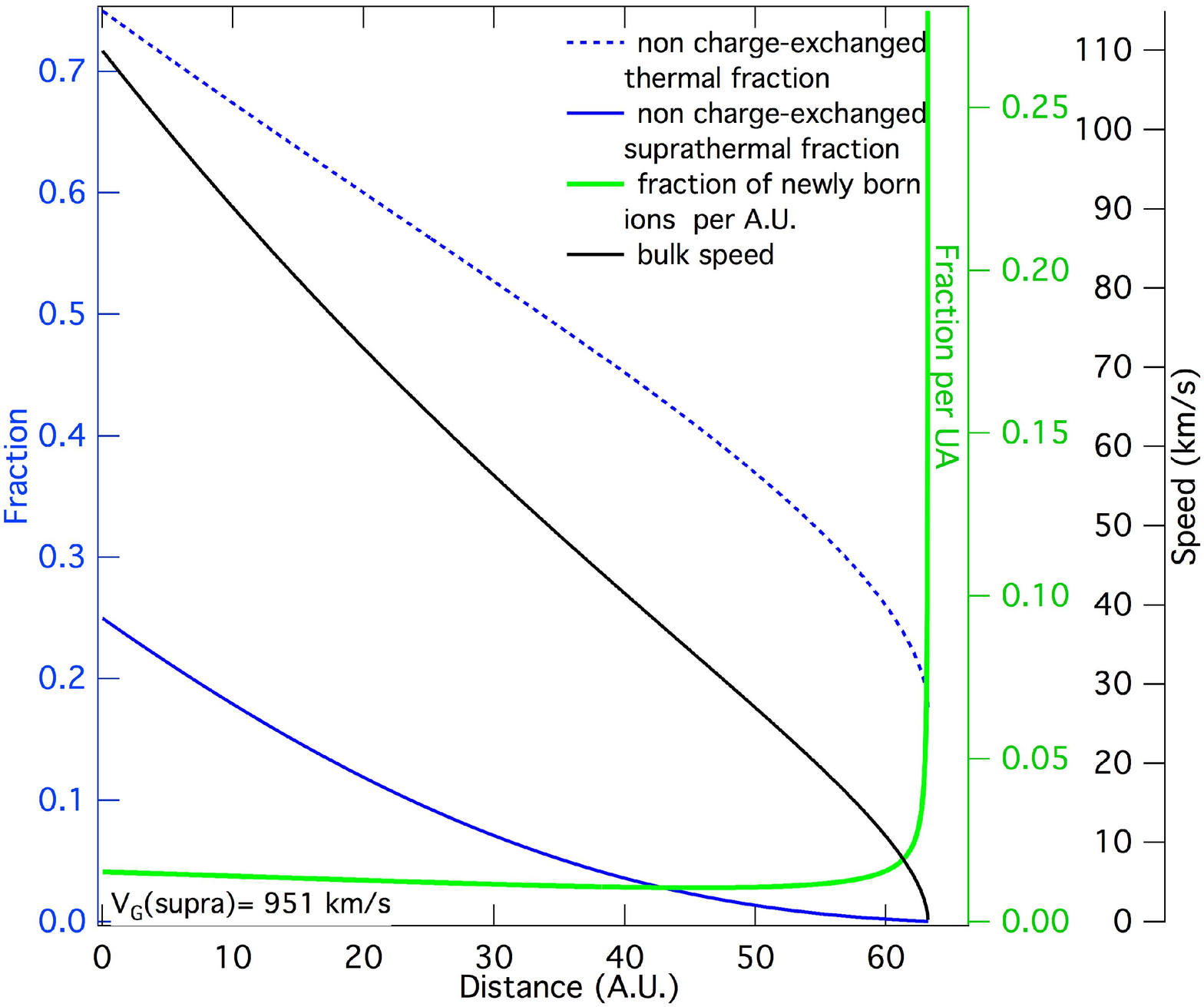}
\includegraphics[width=0.95\linewidth,height=7cm]{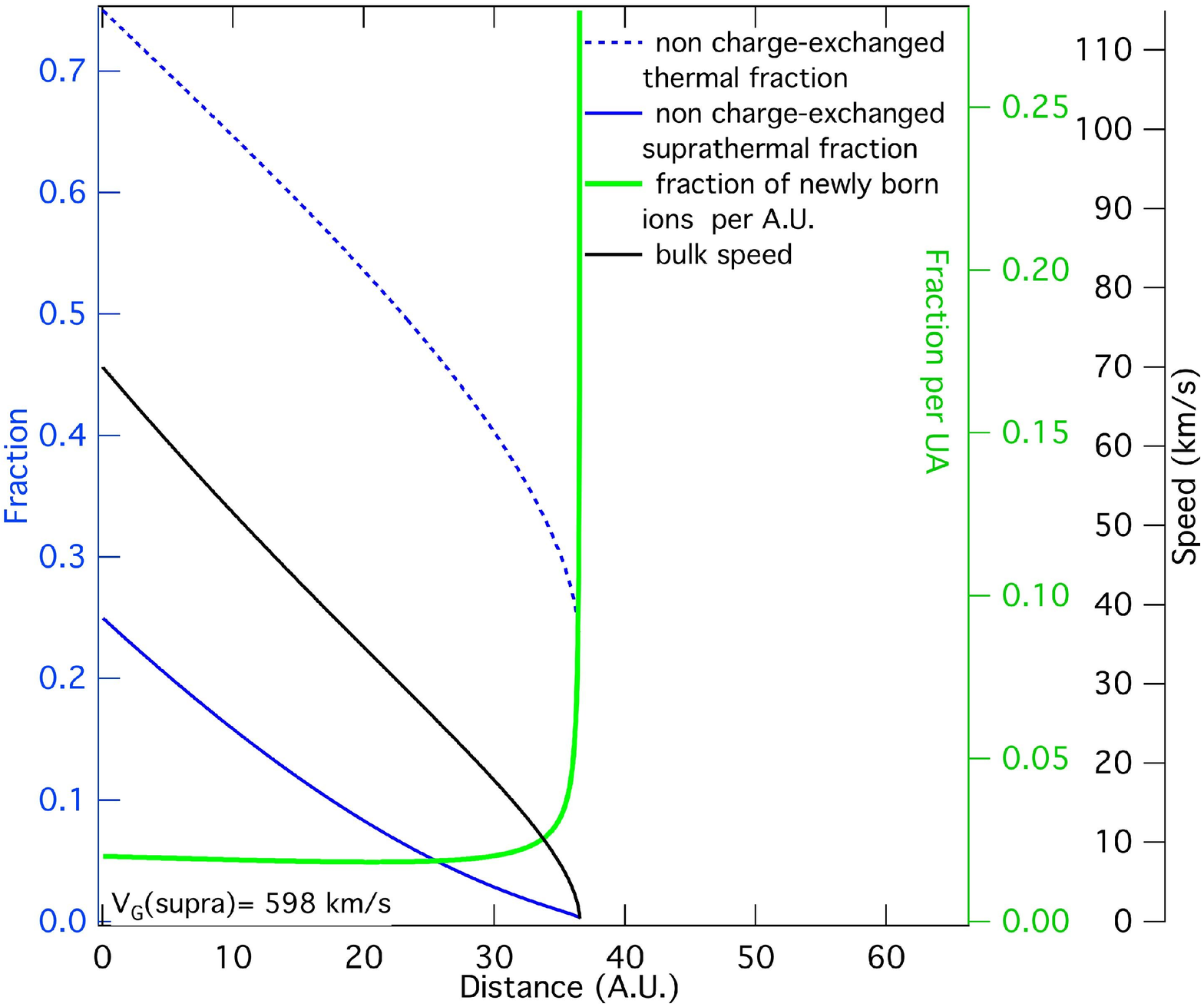}
  \caption{Top: Example of evolution of a flow undergoing CE reactions only. Energetic ions (dashed blue curve) are experiencing the largest amount of CEs due to their large individual speed. After 40 UA they have almost all been trans-charged, for only about half of the ambient ions (blue curve). When the flow is decelerated down to about 5-10 km.s$^{-1}$ (black curve), the CE rate  increases exponentially and the fraction of particles reacting with the neutrals becomes a significant fraction of the plasma, implying a very strong mixing of the two gases through CE (green curve). Here V=110 km.s$^{-1}$,  T=3 10$^{5}$ K, A$_{PUI}$=0.25, n$_{H}$=0.21 cm$^{-3}$, and V$_{H}$= 26.3 km.s$^{-1}$. Bottom: same figure for V=70 km.s$^{-1}$.}
\label{Figdecelhigh}%
\end{figure}

\subsection{Simplified model of CE reactions}
To illustrate the role of CE in the inner heliosheath in low bulk speed conditions, we consider an extremely simplified one-D model of the heliosheath flow in the upwind region. The flow is assumed to be composed of protons and electrons and two initial proton populations, thermal and suprathermal, and its evolution is under the unique influence of charge exchange reactions of the protons with a counter-streaming H atom flow. For simplicity we assume that ions of one population have the same modulus of individual speed in the plasma bulk frame. Each CE reaction corresponds to the replacement of an ion (thermal or suprathermal) by a newly born ion initially at rest in the neutral gas, subsequently accelerated by the motional electric field. We assume that the acceleration is immediate and that there is no energy diffusion at the timescale considered. We also assume that for all newly born ions their final individual speed is equal to the relative velocity between the former neutral and the bulk plasma, as would  be the case if magnetic field lines were everywhere perfectly perpendicular to this relative motion, which is a reasonable assumption in the upwind inner heliosheath, where the field lines are close to ortho-radial and the flow close to radial. Each CE reaction results in a momentum loss of the flow to compensate for the acceleration of the newly formed pickup ion. Although it is not a dominant phenomenon, the newly created neutrals remove a fraction of the flow internal energy, and to take it into account we compute the kinetic energy loss for each individual CE and accordingly decrease the temperature of the thermal fraction of the wind. We also assume that the interstellar neutral flow is continually fully replenished, which is a reasonable assumption to start with, since the flux of neutral H atoms is by far superior to the solar wind flux in the upwind heliosheath. We do not consider any gas or magnetic pressure gradient and neglect electron conduction. There is no exchange between the neutral gas and the wind flow other than through the CE reactions. This simplistic representation is far from actual 3D models but serves only to illustrate the effect of low bulk speeds, of the difference between CE reactions of the two ion populations, and the potential consequences in response to the heliosheath initial characteristics.

The flow, composed of $N$ ions with individual speed $V_{p}$ launched at a bulk speed $V$ is followed at increasing distance $r$. $N=\int{dn(r,V_{p})}$ is constant along the flow since CE reactions maintain the number of protons.  A fraction $A_{PUI}$ of the ions is suprathermal and made of former PUI's with an average speed $V_{p}$= $V_{PUI}$ and the rest is the thermal fraction $(1-A_{PUI})$ with the corresponding individual speed $V_{p}$ being the thermal speed $V_{T}$.  During the ion progression at increasing distance $r$ the bulk speed evolution $V(r)$,  the individual speed distribution $n(r,V_{p})$ and the temperature $T$ are computed according to the CE reactions of the various particles. The neutral flow has a volume density $n_{H}$ and a speed $V_{H}$ (here $V_{H}$ is the absolute value) in a direction opposite to the SW bulk flow. During the time interval that corresponds to the distance step $dr$ of the bulk flow, an ion with an individual speed $V_{p}$ is moving along an helicoidal trajectory whose length does depend only on $V_{p}$. It does not depend on the pitch-angle, even if the actual trajectory depends on the projections of $V_{p}$ onto the magnetic field direction and the orthogonal plane. In this simple 1D model, we do not consider any lateral diffusion along the magnetic field lines and restrict to the average CE rates along the bulk flow, since in 1D any ion drifting perpendicularly to the bulk flow  is replaced by an equivalent ion from the sides. We will discuss departures from this simple scheme below. The CE reaction rate is computed according to the trajectory of each type of ion. For a bulk plasma traveled distance $dr$, the decrease of particles with an individual speed $V_{p}$ under the effect of CE reactions  is:

\begin{equation}
\begin{array}{l}
dn(r,V_{p})= -n(r,V_{p}) * n(H) * \sigma(V_{rel}) * \frac{L(hel)}{L(str)} * dr\\
= -n(r,V_{p}) * n(H) * \sigma(V_{rel}) * \frac{\sqrt{V_{p}^{2}+V^{2}}}{V} * dr
\end{array}
\end{equation}

where $V_{rel}$ is the relative velocity between the ion and the neutral, $\sigma$ is the cross-section for the CE reaction between H atoms and protons, that depends on  $V_{rel}$. 
If $V_{p}^{par}$ and $V_{p}^{perp}$ are the projections of the individual speed onto the bulk motion direction and the orthogonal plane, then $V_{rel}= \sqrt{(V_{p}^{par}+V_{H})^{2}+(V_{p}^{perp})^{2}}=\sqrt{(V_{p}^{2}+V_{H}^{2}+2V_{p}^{par}.V_{H})}$. To simplify the computation we replaced this expression by its average value over the angles with respect to the bulk flow $\sqrt{(V_{p}^{2}+V_{H})^{2}}$. This simplification is justified by the fact that the cross-section does not vary by large factors. $L(hel)/L(str)$ is the ratio between the helicoidal and straight paths for the individual particle of individual speed $V_{p}$, equal to the ratio between the instantaneous speed $\sqrt{V_{p}^{2}+V^{2}}$ and the bulk speed $V$. The length of the helicoidal curve is the same for any pitch-angle of the particle since it does depend on the individual speed only. We note however that the guiding center traveled distance of particles with pitch-angle 90$^{\circ}$ are much smaller than those lengths $L(hel)$, as  in Fig \ref{helices}. For suprathermal ions the high value of this ratio results in enhanced charge-exchange and more generally when the bulk flow is slow the number of CE reactions per traveled distance of the bulk may increase very strongly, especially if newly born ions also have large individual speeds (see fig \ref{helices}). The decrease of the CE cross-section with increasing ion-neutral relative speed is too weak to counterbalance the effect of the high speed ion trajectories, \textbf{for the typical suprathermal ion speeds we consider.}

Each individual CE reaction induces a deceleration of the flow to compensate for the acceleration of the newly born ion from its initial motion at $V_{H}$ (directed against the wind flow) to the bulk flow speed $V$. The resulting effect of CE reactions for ions of all gyrospeeds results in a global deceleration:

\begin{equation}
N * dV(r) = - \int{[dn(r,V_{p}) } * (V_{H} + V(r))]
\end{equation}

For each step $r$ the new individual speed distribution is computed according to the replacement of all charge-exchanged ions by new ions whose individual speed have the value $V_{H} + V(r)$, that depends on the flow speed $V(r)$ and no longer on the initial values of $V_{p}$. Finally, assuming that the neutral particles all escape, the CE reactions result in an internal energy loss of the wind we can compute as the difference between the final and initial individual speeds of the ion. We take into account this energy loss by assuming a corresponding decrease of the thermal speed of the ambient particles.
\begin{equation}
N  d(V_{th}(r)^{2}) = \int{dn(r,V_{p}) *  [(V_{H} + V(r))^{2} -V_{p}^{2})]}
\end{equation}

Tests have shown that the effect of the latter cooling on the bulk speed evolution is in most cases very small. On the contrary, the momentum loss due the CE reactions induces a positive feedback mechanism: as a matter of fact, as schematically illustrated in Fig. \ref{helices}, a bulk speed decrease amplifies the difference between the helicoidal trajectory of all particles and the bulk motion. As a result, the number of CE reactions per unit distance traveled by the flow is increasing. 


We performed a simple step by step integration of the above equations, using distance steps of 0.02 A.U. The individual speed distribution is discretized into a large number of velocity bins (we used 10,000 bins). We use the H-H$^{+}$ charge-exchange cross-section of \cite{lindsay05}. The initial flow parameters are varied around values that are based on the V1 flow  speed determinations of \cite{decker12,krimigis11} based on the Compton-Getty effect and direct Voyager 2 (V2) measurements of \cite{richardson08,richardson11} that provided the main characteristics of the flow beyond the termination shock (TS) at V2: the post-shock temperature and flow speed at V2 were measured to be of the order 1-2 10$^{5}$ K and 100 km.s$^{-1}$ respectively. This corresponds to a thermal speed of the order of 40-55 km.s$^{-1}$. The post-shock flow speed at V1, of the order of 100 km.s$^{-1}$  after the shock, has surprisingly decreased very rapidly down to 50-70 km.s$^{-1}$. We use for the fraction of PUI's formed in the supersonic solar wind and reaching the shock a value of 25\%. The PUI fraction may be derived from global models and solar and interstellar data, but has also been more directly derived from the SW deceleration measured along the V2 trajectory. More specifically, \cite{richardson08_decel} derived a fraction of the order of 17\% when Voyager 2 was at 78 AU. The PUI fraction at the V1 termination shock crossing location is expected to be higher than this values for two reasons: the larger distance on one hand (94 instead of 78 AU), and the smaller angle to the neutral flow on the other hand, because the neutral H flux is at any distance the highest along the upwind axis and ionization are strong in the inner regions of the heliosphere. Using a simple scaling for the former effect alone implies a PUI fraction at V1 larger than 20.5\% (the value obtained if neglecting the H ionization cavity). The second is difficult to determine but we estimate it to potentially increase the PUI fraction further by 10\%. A shown by tests, using 21-22 \% instead of 25\% does not change our main conclusions.

If, as generally accepted now, former pick-up ions (PUIs) capture most of the flow kinetic energy \citep{zank96,richardson08} at the termination shock and
 become a suprathermal population, their average energy can be estimated from the pre-and post-shock flow parameters. Here we simplify largely the computation by assuming they all gather about the same amount of energy at the shock, and compute the corresponding individual speed as a function of the downstream flow speed V, downstream temperature T and relative fraction of those PUI's A$_{PUI}$. As additional simplifying assumptions we consider that prior to the shock the ambient ions thermal speed is negligible, that all PUI's have a gyrospeed equal to the bulk speed, thus carry about twice the kinetic energy of ambient ions, and that the pre-shock speed V$_{UP}$ is four times the downstream speed V. The observed shock at V2 is weaker, however there is a foreshock speed decrease very likely due to the presence of the PUIs, showing that they start to be energized there. Neglecting the electrons, the individual speed V$_{PUI}$ of a suprathermal ion is derived from:

 \begin{equation}
 \begin{array}{l}
 V_{UP}= 4 * V\\
 2* Ec_{UP}/m_{p}=(1-A_{PUI}) * V_{UP}^{2} + 2 * A_{PUI} * V_{UP}^{2}\\
 =V ^{2} + (1-A_{PUI}) * V_{T}^{2} + A_{PUI} * V_{PUI}^{2} \\
 \end{array}
 \end{equation}
The resulting individual speeds are on the order of  500-1000 km.s$^{-1}$ depending on the post-shock flow speed.
 

\subsection{Results}
Fig \ref{Figdecelhigh} shows two examples of the evolution of such an idealized SW flow affected only by CE reactions with a counter-streaming neutral H flow. 
The initial speeds considered are 110 and 70 km.s$^{-1}$ respectively and the temperature is 3 10$^{5}$K. 70 km.s$^{-1}$, a relatively small post-shock speed, is chosen because this is the radial component of the SW flow at V1 in 2010-2011 measured by \cite{decker12}, $\simeq$ 10 A.U. downstream from the shock only. We have chose a temperature slightly above the one measured at V2, because we may expect a stronger TS shock at V1. It can be seen in Fig. \ref{Figdecelhigh} that in both cases energetic ions undergo the largest number of CEs due to their large speed, and are neutralized far in advance of the ambient ions. When the flow is decelerated down to about 5-10 km.s$^{-1}$ the CE rate increases markedly since the individual velocity of most ions exceeds the bulk speed. As a matter of fact even the newly born ions are accelerated to speeds that are of the order of 30-35 km.s$^{-1}$, the relative velocity between the inflowing atoms and the outflowing plasma. Fig. \ref{Figdecelhigh} displays the fraction of the population that is charge-exchanged per traveled A.U., and shows that this fraction becomes extremely large when the bulk flow decreases below 10-5 km.s$^{-1}$. We suggest at this point that such an amplification mechanism may theoretically produce a  stagnation region where a large number of newly born pickup ions are gradually mixed with the solar wind. Evidently this simplistic 1D model breaks down at very low bulk speeds as mass conservation in 1D implies an unrealistically high density, and also when gyration radii become larger than the characteristic length. The actual flow geometry as well as pressure balance must be considered to go one step further. Nonetheless, this crude model shows that when solar wind speed variability and non-stationarity result in a low speed in the upwind region of the heliosheath, CE reactions may become important and amplify the deceleration. We note that the distance of deceleration under the unique effect of CE is very different for the two cases considered in Fig. \ref{Figdecelhigh}: on the order of 60 AU for V=110 km.s$^{-1}$ and on the order of 35 AU for V=70 km.s$^{-1}$. Not shown here is the result obtained for  a post-shock flow of 200 km.s$^{-1}$ characteristic of the high speed solar wind, namely 125 A.U., far above the thickness achieved due to gas and magnetic pressure gradients in global models. This shows that for the high speed wind the influence of the CE deceleration is probably not so important, despite the large speeds of the corresponding PUI's and the suprathermal heliosheath particles they become, simply because the flow speed remains high enough. In the considered models, the influence of the suprathermal fraction is significant, especially for the first phase of the deceleration. For an initial bulk speed of 110 km.s$^{-1}$ the absence of suprathermal increases the deceleration distance by 15 A.U.

\begin{figure}
\centering
\includegraphics[width=\linewidth]{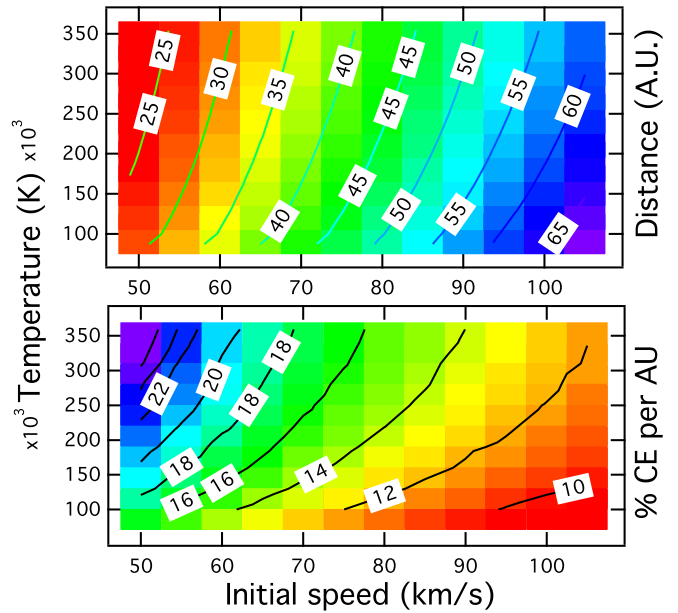}
   \caption{Top: Distance of deceleration by CE momentum transfer down to 1 kms$^{-1}$ as a function of the initial temperature and speed of the flow. Distances are in A.U. (color scale and iso-contours). Other parameters are as in Fig \ref{Figdecelhigh}. Bottom: Fraction of ions experiencing charge-exchange per A.U. traveled distance of the bulk flow, computed where the bulk speed is reduced to 1 km.s$^{-1}$. Fractions are in percent per A.U. (color scale and iso-contour).}
 \label{Figdecelparam}%
\end{figure}

Fig \ref{Figdecelparam} is a limited parametric study and shows the influence of the initial bulk speed and temperature on the deceleration. It displays the length corresponding to the complete deceleration of the flow down to 1 km.s$^{-1}$, and the fraction of ions experiencing CE per traveled A.U. at the threshold of 1 km.s$^{-1}$. We see that: (i) the deceleration of the wind by the unique effect of charge-exchange reactions with the neutral flow may occur after distances  as small as 25 A.U. if the wind bulk speed is smaller than 70 km.s$^{-1}$, for reasonable values of the other parameters. For comparison, the distance traveled by V1 in the heliosheath  between 2004 and 2013 is about 30 A.U. Of particular interest is the fact that, as mentioned above, after the TS crossing V1 experienced a particularly slow wind, certainly due to non-stationary effects \citep{jokipii05,pogorelov12}. The radial speed remained very low and often below 50 km.s$^{-1}$ during the first two or three years, before starting a gradual and fast decrease. We note a similarity with the simulated deceleration. (ii) As expected, Fig \ref{Figdecelparam} shows that the distance of deceleration decreases strongly with the initial speed. However, Fig. \ref{Figdecelhigh} shows that there is no linearity, as can be derived e.g. from the two cases shown in this figure. A flow with initial speed 110 km.s$^{-1}$ looses 40 km.s$^{-1}$ over 41 A.U. to reach 70 km.s$^{-1}$. Over the same distance a flow with initial speed 70 km.s$^{-1}$ has been decelerated to zero over  37 A.U. This shows that if adjacent flows have different speeds and similar composition and temperature, the faster flow will be less decelerated than the slower, increasing the speed difference. This holds for consecutive streams: a fast stream chasing a slower one should decelerate less and thus overtake it more rapidly, while the gap between a fast and a slow stream lagging behind should increase.   (iii) The distance of deceleration also decreases with increasing temperature through  the ion individual speed increase.

\subsection{Charge-exchange along the V1 and V2 trajectories}
If the bulk speed decreases to reach values as low as 5-10 km.s$^{-1}$ the number of \textbf{CE} reactions increases very rapidly, a self-amplification mechanism for the deceleration and the mixing of the plasma with the neutrals. Such a self-amplified deceleration can occur:

-(i) where the solar wind bulk flow is opposite to the interstellar flow, for the momentum exchange to counteract the global motion. At large angles the flow starts to deviate before reaching low speeds, precluding any amplification.

-(ii) when and where the post-shock speed is not too high, i.e. it excludes the high-speed wind.

-(iii) as seen above, after particular episodes of strong variability and non-stationary low speed winds.

As far as (i) and (iii) are concerned the situations were different for V1 and V2. First, the angle with the wind axis is larger by 20$^{\circ}$ for V2. Second, there is no low speed episode for V2 such as those having occurred at V1, and after a short distance from the TS the wind along the V2 trajectory rapidly started to deviate, while still keeping a high speed. More generally solar variability effects were much weaker at V2, as shown by the absence of strong fluctuations in the measured speed nor in density, at variance with the strong fluctuations at V1 measured by LECP. Solar cycle effects have been shown by \cite{pogorelov12} to explain those different situations: V1 entered the heliosheath at a time of strong solar wind pressure decrease during the post-solar maximum, which may explain sunward motions of the TS and the peculiarly low speeds experienced by Voyager 1. As seen above this creates particularly favorable conditions for amplified CE deceleration, potentially helping to reach stagnation after a short distance on the order of 30 A.U., while this is the opposite for V2.

\section {Conclusions and discussion}

In a first part of this article we have studied in detail the temporal evolution of the UVS detector background since 1992, and shown that it is essentially due to Galactic cosmic rays with energies of the order of 300 MeV. A minor contribution from the particles created by the RTG is not precluded, however it has no impact on the temporal evolution of the background or on our conclusions. 
Using data from the CRS and LECP instruments and assuming they are detecting mostly protons, we showed that the UVS detector background is due to protons slightly more energetic than the ones detected by the CRS-IPGH detector, but that in contrast with IPGH it is significantly sensitive to particle flux anisotropies. Using the UVS data in combination with the other cosmic ray instruments, we showed that the way the GCR proton spectral slope around 300 MeV has evolved precludes with a high probability a future increase of the flux by more than a few per cent. In other words, the demodulation is completed or almost completed. This implies that the thickness of the heliosheath beyond the S/C location in August 2012 is very small, if not null.

Such a conclusion is in line with other results based on Voyager 1 data suggestive of an entrance in the interstellar medium: (i) the heliosheath energetic particle disappearance \citep{krimigis13}; (ii) the GCR abrupt increases and the new spectra \citep{stone13,webber13,webber13b} (iii) the solar wind stagnation since 2010 \citep{decker12,krimigis11}; (iv) the comparisons between the Cassini, SOHO/HSTOF and IBEX data on one hand and Voyager particle data on the other  hand that provided estimates of the heliosheath thickness  in the Voyager 1 direction (21 $\pm6$ \citep{hsieh10,czecho12} and 31(+31,-18) AU \citep{roelof12}). 
(v) the constancy of the IPGH proton flux (plus UVS, if corrected for anisotropies), for more than a year now. This situation never happened since at least 1992; (vi) the recent measurements of plasma oscillations that imply a high density at V1 \citep{gurnett13}.

All the above measurements, if interpreted as entrance in the ISM, imply a heliosheath that is thinner than heliospheric models predict. In a second part of this article we have investigated the potential deceleration due to charge-exchange with counter-streaming interstellar H atoms in non-stationary conditions,  and subsequent shrinking of the heliosheath. 
With the help of an extremely simple model we have compared the effects of the CE on post-shock flows for various conditions and showed that they are significant in case of  post-shock solar wind low speed episodes occurring on the upwind side. We suggested that the deceleration generated by charge-exchange reactions in such cases may produce a positive feedback and significantly help in reducing the thickness of the heliosheath. In case of very low speeds such as those measured at V1 \citep{decker12,krimigis11} we suggest that the contact discontinuity between the solar and interstellar plasmas may be replaced by a stagnation layer mixing the two gases, in which  the transition is ensured by CE reactions, somewhat similarly to the hot-cool gas interface modeled by \cite{provornikova11}. Favorable conditions were present along the V1 trajectory after 2004, while for V2, that travels at a larger angle with the heliosphere axis and did not experience low-speed episodes, the CE deceleration remained negligible. We note that in a situation like the one described in Fig \ref{Figdecelhigh}, where a very large fraction of the flow is continuously experiencing CE, the conservation of the magnetic flux may not be ensured. This could explain why the magnetic flux was conserved for V2, and decreased significantly for V1 \citep{richardson13}. We  suggest that due to the large rate of CE reactions, the  flux of simultaneously newly born neutrals  (Fig \ref{Figdecelhigh}) is significant and those neutrals may again charge-exchange with piling-up upstream interstellar ions beyond the heliopause. The whole ensemble of reactions may lead to the disappearance of a distinct boundary and the formation of a stagnation layer of mixed solar and interstellar ions. We note that \cite{provornikova11,provornikova12} already modeled such a transition layer between hot and cool gases and showed that CE in this layer may play the role of electron conduction.
We finally speculate that due to enhanced CE reactions the upwind heliopause may present a concave region elongated in the equatorial plane where slow solar wind dominates, which could influence the plasma flows inside and outside. If the interstellar flow is guided into and out of this carved region instead of being diverted according to the classical scheme, this may help explaining why the magnetic field has kept a direction that is close to the Parker spiral.

The 1D assumption we used is questionable for particles that may diffuse rapidly away from  the upwind region, if they are not being replaced by equivalent particles from the sides. Suprathermal particles are particularly concerned. Their diffusion into the flanks of the heliosheath is however reduced if the large pitch-angle values they possess after the TS crossing is maintained, i.e.  isotropization is not too fast. On the other hand two mechanisms help maintaining large pitch-angles in the upwind inner heliosheath. Firstly, if  the flow decelerates magnetic field lines are compressed, the field intensity increases, and by conservation of the magnetic moment pitch-angle also increase (we still restrict to perpendicular magnetic field). Indeed, the magnetic field at V1 has increased by a factor of three between the post-TS region and 2013 \citep{richardson13,burlaga13}. Second, newly created and accelerated pickup ions are injected mainly at 90$^{\circ}$. More realistic models should help quantifying those effects. As a matter of fact  the present study does not allow to reach definite conclusions, but it suggests that CE reactions of suprathermal and thermal  ions may be important in non-stationary conditions in the upwind heliosheath, and that this potential amplification mechanism deserves further investigations. In particular the creation of a collapsed charge-exchange layer, as well as a potential irreversibility of this collapse deserve more studies.

Future Voyager 1 field, particle and hopefully wave measurements should continue to bring crucial information in the next years. The evolution of the Lyman-alpha diffuse emission should also bring crucial additional constraints as far as the suggested mixing layer is concerned, since the emission is strongly dependent on the neutral gas density distribution and in particular the thickness and height of the neutral H \textit{wall}, itself strongly linked to the interstellar plasma \textit{wall}. Finally, the end of the GCR modulation may be confirmed (or not) by the analysis of the diffuse gamma rays recorded by the NASA Fermi satellite. Low energy gamma-rays are mainly due to the interaction of the galactic cosmic rays with the interstellar medium and radiation field in the vicinity of the Sun. The Fermi data analysis should provide estimates of the GCR fluxes down to 1 GeV. If the Voyager 1  ($\leq$ 0.45 GeV)  GCR spectrum in September 2012 is the true interstellar spectrum, then the Fermi spectrum above 1 Gev should be in line with the CRS and LECP extrapolated at higher energies.

\begin{acknowledgements}
\textbf{We thank our anonymous referee for his critical analysis that resulted in a significant improvement of the second part of this article.}This work has been funded by CNRS and CNES grants. We are greatly indebted to Terry Forrester from LPL (Tucson) for his long-term involvement in the UVS data reduction and calibration. 
The Voyager CRS data were obtained from: http://voyager.gsfc.nasa.gov
\end{acknowledgements}


\end{document}